\documentclass[aps,twocolumn,showpacs]{revtex4}

\usepackage{graphicx}

\begin{document}

\def\bsigma{\mbox{\boldmath $\sigma$}}
\def\bomega{\mbox{\boldmath $\omega$}}

      \title{Hot carriers in an intrinsic graphene}
      \author{O.G. Balev}
      \affiliation{Departamento de Fisica, Universidade Federal do Amazonas,
       Manaus, 69077-000, Brazil}
      \author{F.T. Vasko}
      \email{ftvasko@yahoo.com}
      \affiliation{Institute of Semiconductor
      Physics, NAS of Ukraine, Pr. Nauki 41, Kiev, 03028, Ukraine}
      \author{V. Ryzhii}
      \affiliation{University of Aizu, Ikki-machi, Aizu-Wakamatsu
      965-8580, Japan}
      \affiliation{Japan Science and Technology Agency,
      CREST, Tokyo 107-0075, Japan}
\date{\today}

\begin{abstract}
Heating of carriers in an intrinsic graphene under dc electric
field is considered taking into account the intraband energy
relaxation due to acoustic phonon scattering and the interband
generation-recombination transitions due to thermal radiation. The
distribution of nonequilibrium carriers is obtained for the cases
when the intercarrier scattering is unessential and when the
carrier-carrier Coulomb scattering effectively establishes the
quasiequilibrium distribution with the temperature and the density
of carriers that are determined by the balance equations. Because
of an interplay between weak energy relaxation and
generation-recombination processes a very low threshold of
nonlinear response takes place. The nonlinear current-voltage
characteristics are calculated for the case of the momentum
relaxation caused by the elastic scattering.  Obtained
current-voltage characteristics show low threshold of nonlinear
behavior and appearance of the second ohmic region, for strong
fields.
\end{abstract}


\maketitle

\section{Introduction}
Active study of graphene, taking place in recent years, is stimulated both
by unusual physical properties of this gapless and massless semiconductor
(see discussion and references in the review \cite{1}) and by the need to
study the possibilities of applications in electronics, see discussion in
\cite{2}. For such applications study of a response
      of nonequilibrium carries is important and very much needed. Both
      experimental and theoretical studies of the nonequilibrium
      electron-hole pairs in graphene under interband photoexsitation
      have been already performed, see \cite{3} and \cite{4,5} (and
      references therein), respectively. Experimental investigations
      of the carries heating in graphene by dc electric field were
      carried out in relation with demonstrations of graphene
      field-effect transistor (see \cite{6} and references therein).
      Essential heating apparently had taken place in experiments on
      current-induced cleaning of graphene \cite{7}. In respect to
      theoretical treatments of the problem, it was investigated the
      rate of energy relaxation of nonequilibrium carries, both
      analytically \cite{8} and numerically,\cite{9}  and also it was
      discussed a derivation of kinetic equation for a strong field
      \cite{10}. However, to the best of our knowledge, the basic problem
      of carries heating by dc electric field have not been considered
      so far and a systematic investigation of the non-linear
current-voltage characteristics was not performed yet.

      In this paper, we study a heating of electron-hole pairs in an
      intrinsic graphene under dc electric field. To describe heating it
      is taken into account both increase of the energy of carries in the
      field, $\bf E$, and the following relaxation processes: (1) damping
      of the momentum due to elastic scattering on structural
      disorder, that is the most fast process ensuring weak anisotropy
      of the distribution function  \cite{11}, (2) intraband
      quasi-elastic energy relaxation on acoustic phonons, (3)
      generation-recombination  processes due to interband transitions
      induced by thermal radiation, and (4) intercarrier scattering due to
Coulomb interaction. In present study we consider the heating
within two limiting regimes: ($I$) if the Coulomb scattering is
unessential, when it it is possible to neglect by the intercarrier
scattering, and ($II$) the Coulomb-controlled case. In the latter
case the Coulomb scattering imposes the quasiequilibrium
      distribution on carries, with characteristics defined
      by balance equations for the density and the energy. In  present
      study we find the distribution function of carries for these
      regimes and investigate it dependence on applied field and the
      temperature of thermostat formed by phonons and thermal radiation.
      In addition, we analyze current-voltage characteristics and the
      density of nonequilibrium carriers.

      The character of obtained nonlinear response is determined by a few
      peculiarities of the considered model. First, the velocity of
      carries does not increase with the energy as it is equal $v_W{\bf
      p}/p$, where $v_W=10^8$ cm/s is the characteristic velocity of
      charged neutrino-like particles, $\bf p$ is the 2D momentum.
Second, the interband transitions are effectively excited by thermal radiation (the
matrix element of transition is $\propto v_W$) they not only change a
carrier concentration but also participate in the energy relaxation. As
a result, the nonequilibrium distribution is formed due to interplay
between weak energy relaxation and generation-recombination processes (i.e.,
between interactions with phonon and photon thermostats).
Third, the rate of energy relaxation sharply increases as $p$
      grows, where weak electron-phonon interaction takes place for slow
      carriers. Therefore for scattering on short-range defects, when
      the relaxation of momentum also grows with the energy, it is
      realized a sub-linear current-voltage characteristic with essential
      nonlinearity for weak fields while as field grows nonlinearity
becomes weak.  If the momentum relaxation becomes ineffective for high
energy carriers (because of a finite-range disorder) a super-linear increase
of  current takes place. In the region of strong fields a linear dependence
of current from the field is realized (the second ohmic region of current-voltage
characteristic) where the effective conductivity is smaller than the conductivity 
of equilibrium carriers for low temperatures, $T$, and the former becomes
larger than the latter as $T$ grows.
Because the intrinsic graphene has maximum resistance (with respect
to gate voltage), here the Joule heating is least effective so maximal electric
field can be applied.

      The paper is organized in the following way. The basic equations
      describing the heating of carriers in an intrinsic graphene under
      dc electric field are presented in Sec. II. In Sec. III we examine
      the symmetric distribution function of nonequilibrium
      electron-hole pairs for the cases $I$ and ($II$), see above. The
      current-voltage characteristics are analyzed in Sec. IV. The
      concluding remarks and discussion of the assumptions used are
      given in the last section. Appendix contains the evaluation of the
      quasiclassical kinetic equation for electron-hole pairs under a strong dc
      electric field.

      \section{Basic equations}

Nonequilibrium electron-hole pairs in the intrinsic graphene are described
by coinciding distribution functions $f_{e,h{\bf p}}\equiv f_{\bf p}$
as their energy spectra are symmetric and processes of scattering there are
identical (see Appendix). Therefore instead of system of kinetic equations for
$f_{e{\bf p}}$ and $f_{h{\bf p}}$ it is possible to consider the single kinetic
equation as
      \begin{equation}
      e{\bf E}\cdot\frac{\partial f_{\bf p}}{\partial {\bf p}}=
      \sum_jJ_j\{f|{\bf p}\} ,
      \label{1}
      \end{equation}
where $\bf E$ is a strong electric field and the classical form of
      the field term is substantiated in Appendix. Here ${\bf p}$ is the
      2D momentum and $J_j\{f|{\bf p}\}$ is the collision integral for
      the $j$th scattering mechanism, with index $j=D,~LA, ~R$, and $C$
      correspond to the static disorder ($D$), the acoustic phonon
      scattering ($LA$), the radiative-induced interband transitions
      ($R$), and the carrier-carrier scattering ($C$), respectively.
      Integrals of elastic scattering there were considered earlier,
      see \cite{11, 12} and references therein, $J_{LA}\{f|{\bf p}\}$
      and $J_{R}\{f|{\bf p}\}$ were evaluated in \cite{5}, and Coulomb
      scattering was considered in \cite{13}.  For the distribution
      function defined by Eq. (\ref{1}), the concentration balance equation is
      given as
      \begin{equation}
      \frac{4}{L^2}\sum_{\bf p}J_R\{f|{\bf p}\} =0 ,
      \label{2}
      \end{equation}
      because interband transitions are forbidden not only for elastic
      scattering but also for Coulomb scattering, due to symmetry of the
      energy spectrum \cite{14}, and for phonon scattering, due to $s\ll
      v_W$; here $s$ is the velocity of sound. In Eq. (\ref{2}) the factor 4
      takes into account spin and valley degeneracies, $L^2$ is the
      normalization area.

      Taking into account that electrons and holes equally contribute
      to the current density, ${\bf I}$,  one obtains
      \begin{equation}
      {\bf I}=\frac{8ev_{\ss W}}{L^2}\sum_{\bf p}\frac{\bf p}{p}\Delta
      f_{\bf p} ,
      \label{3}
      \end{equation}
where the asymmetric part of distribution function, $\Delta f_{\bf p}$, is
separated from the symmetric one, $f_p$, by using the relation $f_{\bf p}=f_p+
\Delta f_{\bf p}$. Last contribution is small if usual condition $|e{\bf E}|
\tau_p^{\ss (m)}\ll p$ is satisfied. \cite{15} Here $\tau_p^{\ss (m)}$ is the
momentum relaxation time and $\Delta f_{\bf p}$ is given as
      \begin{equation}
      \Delta f_{\bf p}=\frac{(e\mathbf{E}\cdot \mathbf{p})}p\tau _p^{\ss
      (m)}\left( -\frac{df_p}{dp}\right).
      \label{4}
      \end{equation}
For the case of a random potential $U_{\bf x}$ characterized by the correlation
      function $\langle U_{\bf x}U_{\bf x'}\rangle\equiv\overline{U_d}^2\exp\{
      -[({\bf x}-{\bf x'})/l_c]^2\}$ with the averaged energy $\overline{U_d}$
      and
      the correlation length $l_c$, one obtains \cite{11}
      \begin{equation}
      \frac{1}{\tau_p^{\ss (m)}}=\frac{v_dp}{\hbar}\Psi
      \left(\frac{pl_c}{\hbar}\right) , ~~~~ \Psi
      (z)=\frac{e^{-z^2}}{z^2}I_1\left( z^2\right) ,
      \label{5}
      \end{equation}
      where the decreasing with $pl_c/\hbar$ form-factor $\Psi (z)$ is
      written through the modified Bessel function $I_1(x)$ and the
      characteristic velocity $v_d$ is introduced as $v_d=\pi
      \overline{U}_d^2l_c^2/4\hbar ^2v_W$. The conductivity, $\sigma$ is
      determined according to the standard formula ${\bf I}=\sigma{\bf E}$. Using
      the above-introduced relaxation time (5) and Eqs. (3, 4), one transforms the
      conductivity as follows
      \begin{equation}
      \sigma=\frac{e^2}{\pi\hbar}\frac{2v_W}{v_d}\int_0^\infty\frac{dp}
      {\Psi (pl_c/\hbar )} \left( -\frac{df_p}{dp}\right) , \label{6}
      \end{equation}
      where the averaging over angle is performed.

Further, we turn to the averaging over angle (below we symbolize such averaging as
a line over expression) of Eq. (\ref{1}). Neglecting
      the weak contribution of $\Delta f_{\bf p}$ in the right-hand side of
      (\ref{1}), one obtains the kinetic equation for symmetric distribution
      $f_p$ in the following form
      \begin{equation}
      \overline{e{\bf E}\cdot\frac{\partial \Delta f_{\bf p}}{\partial
      {\bf p}}}=\sum_jJ_j \{ f|p\}  .
      \label{7}
      \end{equation}
      Here $J_j \{ f|p\}=\overline{J_j\{ t|{\bf p}\}}$ and summation is
      performed over $j=LA,~R,~C$ because the elastic scattering does not
      affect the symmetric distribution due to the energy conservation.
      Performing the averaging over the angle, we transform the field
      contribution of Eq. (\ref{7}) as follows \cite{16}
      \begin{equation}
      \overline{e{\bf E}\cdot\frac{\partial \Delta f_{\bf p}}{\partial
      {\bf p}}} = \frac{(eE)^2} {{2p}}\frac{d}{{dp}}p\tau_p^{\ss
      (m)}\left( -\frac{df_p}{dp}\right) .
      \label{8}
      \end{equation}
As a result, using the Fokker-Planck form of $J_{LA}\{ f|p\}$ and
$J_{\ss R}\{ f|p\}$ obtained in \cite{5}, we arrive to kinetic equation
      \begin{eqnarray}
      \frac{\nu_p^{(qe)}}{p^2}\frac{d}{dp}\left\{\left[
      p^4+\frac{p^4_E}{2 \Psi (pl_c/\hbar )}\right]
      \frac{df_p}{dp}+\frac{p^4}{p_T}f_p(1-f_p)\right\}  \nonumber \\
      +\nu_p^{\ss (R)}\left[ N_{2p/p_{\ss T}}(1-2f_p)-f_p^2\right]
      +J_{\ss C} \{ f|p\}=0, \label{9}
      \end{eqnarray}
where $N_x=(e^x-1)^{-1}$ is the Planck distribution and $p_T=T/v_W$.
The rates of quasi-elastic energy relaxation, $\nu_p^{\ss (qe)}$, and
of radiative transitions, $\nu_p^{\ss (R)}$, can be presented
in the form \cite{5}
      \begin{eqnarray}
      \nu_p^{\ss (qe)}=\left(\frac{s}{v_W}\right)^2\frac{v_{ac}p}{\hbar}
      , ~~~ v_{ac}=\frac{D^2T} {4\hbar^2\rho_sv_Ws^2} ,\nonumber \\
      \nu_p^{\ss (R)}=\frac{v_rp}{\hbar}, ~~~
      v_r=\frac{{e^2\sqrt\epsilon}}{{\hbar c}}\left( \frac{v_w}{c}
      \right)^2 \frac{8v_W}{3} .
      \label{10}
      \end{eqnarray}
      The characteristic velocities $v_{ac}$ and $v_r$ are introduced
      here in order to separate the linear momentum dependence and
      expressed through the deformation potential and the sheet density
      of graphene, $D$ and $\rho_s$, as well as the dielectric
      permittivity, $\epsilon$. Characteristic momentum, $p_E$, is
      defined in Eq. (\ref{9}) by relation
      \begin{equation}
      p_E^4=\left(\frac{v_{W}}{s}\right)^2\frac{(eE\hbar )^2}{v_{ac}v_d}  ,
      \label{11}
      \end{equation}
so that $p_E\propto\sqrt{E}$. The Coulomb scattering integral $J_C\{ f|p\}$
can be neglected in Eq. (\ref{9}) for the case $I$ therefore here nonequilibrium
distribution $f_p$ is defined by nonlinear differential equation of the second
order.

For other approach, in the case $II$, the Coulomb scattering term in Eq. (\ref{9})
is dominant and it imposes quasiequilibrium distribution
      \begin{equation}
      \widetilde{f}_p=\{\exp [(v_{W}p-\mu )/T_c]+1\}^{-1} ,\label{12}
      \end{equation}
      with effective temperature of carriers $T_c$ and quasichemical
      potential $\mu$. To determine $T_c$ and $\mu$ we will use the
      concentration balance equation (\ref{2}) and the energy balance
      equation. The latter we obtain by summing of Eq. (\ref{7}) over
      $\bf p$ with the energy weight $v_Wp$ as follows
      \begin{equation}
\frac{1}{2}\sigma E^2+\frac{4v_W}{L^2}\sum_{\bf p}p[J_{LA}(f|p)+
J_{R}( f|p)]=0 , \label{13}
      \end{equation}
      where the field term is expressed through the conductivity
      (\ref{6}) using the integration by parts (the factor $1/2$
      is quite understood as the total Joule heat $\sigma E^2$ must to
      be divided equally between electrons and holes). Therefore, the
      two parameters of $\widetilde{f}_p$ are determined by two transcendent
      equations (\ref{2}) and (\ref{13}). Now, the nonlinear conductivity
should be calculated under substitution $\widetilde{f}_p$ into Eq. (6).

      \section{Nonequilibrium distribution}

Below we consider the nonequilibrium distribution obtained from Eq. (\ref{9})
with the Coulomb contribution omitted (case $I$) or from the balance equations
(\ref{2}) and (\ref{13}) (case $II$). Also we present the field and temperature
dependencies of the nonequilibrium sheet concentration.

      \subsection{Weak intercarrier scattering}
We start with consideration of the case $I$, when the carrier-carrier scattering
is ineffective. Omitting $J_C$ in Eq. (\ref{9}) and introducing the dimensionless
momentum $x=p/p_{T}$ and the parameter $\eta =p_{T}l_c/\hbar$, we obtain the nonlinear differential equation
      \begin{eqnarray}
      \frac{d}{dx}\left\{\left[ x^4+\frac{(p_E/p_T)^4}{2 \Psi (x\eta )}\right]
      \frac{df_x}{dx}+x^4f_x(1-f_x)\right\}  \nonumber \\ +\Gamma
      x^2\left(\frac{1-2f_x}{e^{2x}-1}-f_x^2\right) =0. ~~~~~
      \label{15}
      \end{eqnarray}
Here the dimensionless parameter $\Gamma =(v_W/s)^2v_r/v_{ac}$ determines
the ratio of contributions from the thermal radiation and acoustic phonons,
to the energy relaxation. As the boundary conditions for Eq. (\ref{15}) we use
that $x^4(df_x/dx+f_x)_{x\to\infty}=0$, so that $f_x$ must decrease for $x\to\infty$
sufficiently fast, and also the density balance equation (\ref{2}). \cite{15} If to
integrate Eq. (\ref{15}) over $x$ from $0$ to $\infty$, then for finite electric
field besides  Eq. (\ref{2}) we obtain additional, proportional to $(p_E/p_T)^4
(df_x/dx)_{x=0}$, term that, certainly, must be equal to zero. Thus, instead of
Eq. (\ref{2}), the second boundary condition can be used in the following form
$(df_x/dx)_{x\to 0}=0$.
\begin{figure}[ht]
\includegraphics{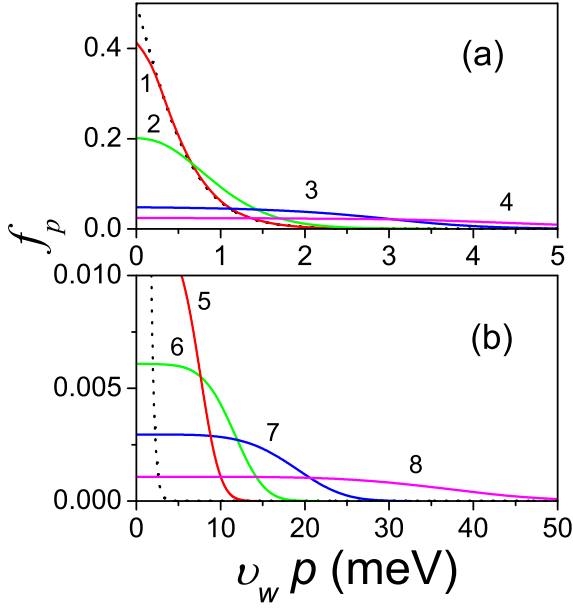}
\caption{(Color online) Determined from Eq. (\ref{15})
distribution functions, at $T$=4.2 K and $l_c=$10 nm: (a) for weak
electric fields $E=$0.1 mV/cm (1), 1 mV/cm (2), 10 mV/cm (3), and
30 mV/cm (4); (b) for strong electric fields $E=$0.1 V/cm (5), 0.3
V/cm (6), 1 V/cm (7), and 5 V/cm (8). Dotted curves correspond to
the equilibrium distribution.}
\end{figure}

The numerical solution of Eq. (\ref{15}) is performed below with
the use of the finite difference method and the iterations over
non-linear contributions, see \cite{17}. Distribution functions
obtained are presented in Figs. 1 and 2 as functions of the energy
for different temperatures, electric fields, and $l_c$. For low
temperature region, two regimes of the modification of a
distribution emerge with the increase of field: weak electric
fields, $(p_E/p_T)^4/(4\Gamma) \ll 1$, and strong electric fields,
$(p_E/p_T)^4/(4\Gamma) \gg 1$.

As it is seen from Figs. 1a and 2 in weak fields there is
essential suppression of the distribution in the region of small
$p <p_T$ (as slow carriers are heated very effectively) and for
large $p>p_T$ distribution remains almost equilibrium one.
Further, as the field becomes strong, Figs. 1 and 2 show that the
carriers are spread over a wide region of energies such that,
e.g., at $T$=4.2 K for $E>$5 V/cm the tail of the distribution
will reach the energies for which the spontaneous emission of the
optical phonons begins ($\sim$90 meV). It is seen, as well in
agreement with Figs. 1 and 2, that with the increase of $T$ the
transition between the weak field and the strong field regimes
occurs at a higher $E \propto T^{2}$. In Figs. 2a - 2c we plot the
distribution functions for $l_{c}=10$nm and $20$nm for different
$E$ and $T$. It is seen that with the increase of $E$ the
influence of a finite $l_{c}$ grows, by making the effect of
electric field on the distribution function more pronounced for
higher $l_c$. However, for weak $E$ the difference between the
solid curve and it dashed counterpart is practically invisible.
Notice, from the solid and, especially, the dashed curves 4 in
Fig. 2c it follows that here interaction with optical phonons can
be important.

\begin{figure}[ht]
\includegraphics{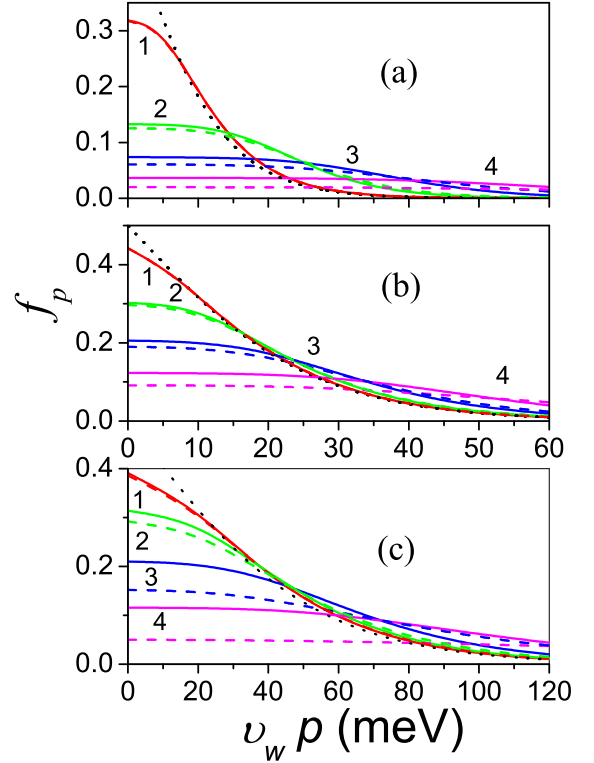}
\caption{(Color online) The same as in Fig. 1 calculated: (a) for
$T$=77 K and $E=$0.14 V/cm (1), 1.4 V/cm (2), 4.4 V/cm (3), and 14
V/cm (4); (b) for $T$=150 K and $E=$74 mV/cm (1), 0.74 V/cm (2),
2.35 V/cm (3), and 7.4 V/cm (4);(c) for $T$=300 K and $E=$1 V/cm
(1), 3 V/cm (2), 10 V/cm (3), and 30 V/cm (4). The solid and the
dashed curves correspond to $l_c=10$ nm and 20 nm, respectively.
Dotted curves correspond to equilibrium distributions.}
\end{figure}

\subsection{Coulomb controlled distribution}
Further, we examine the case $II$, when quasiequilibrium
distribution Eq. (\ref{12}) is determined from the balance
equations, (\ref{2}) and (\ref{13}). Introducing dimensionless
momentum, $y=v_{W}p/T_{c}$, such that $\widetilde{f}_y=[\exp
(y-\mu /T_c)+1]^{-1}$, we rewrite the concentration balance
equation (\ref{2}) in the form
      \begin{equation}
      \int_0^{\infty}dyy^2\left(\frac{1-2\widetilde{f}_y}{e^{2yT_{c}/T}-1}-
      \widetilde{f}_y^2\right) =0 , \label{16}
      \end{equation}
which gives the relation between $\mu /T$ and dimensionless temperature
$T_c/T$ (this relation is not dependent explicitly on the field). Now
the energy balance equation (\ref{13}) is given as
      \begin{eqnarray}
      Q_E-\frac{T_{c}-T}{T}\int_0^{\infty}dyy^4e^{y-\mu
      /T_c}\widetilde{f}_y^2 \nonumber  \\
      +\Gamma\int_0^{\infty}dy y^3 \left(\frac{1-2\widetilde{f}_y}{e^{2yT_{c}/T}-1}-
      \widetilde{f}_y^2\right) =0,  \label{17}
      \end{eqnarray}
where field contribution $Q_E$ is transformed  using Eq. (\ref{6})
      \begin{equation}
      Q_E=\left(\frac{p_E}{T_{c}/v_W}\right)^4\left[\widetilde{f}_{y=0}+
      \frac{\eta_c}{2}\int_0^\infty dy\widetilde{f}_y\Phi (\eta_{c}
      y)\right] .
      \label{18}
      \end{equation}
Here $\eta_c =T_cl_c/\hbar v_W$ and it is introduced the function
$\Phi (z)=-\Psi '(z)/\Psi(z)^2$ and taken into account that $\Psi (0)=1/2$.

In Figs. 3a and 3b calculated from Eqs. (\ref{16}), (\ref{17})
dimensionless effective temperature $T_c/T$ and maximal value of
the distribution function $\tilde{f}_{p=0}=[\exp(-\mu/T_c)+1]^{-1}$,
that determines quasichemical potential, are shown as function of $E$
for different $T$. From
Fig. 3a it is seen that $T_c/T$ grows faster with the increase of
$E$ for larger $l_{c}$; in addition, the grows of dimensionless
effective temperature  becomes faster for smaller $T$. Figure 3b
shows that the characteristic value of distribution, $f_{p=0}$,
decreases faster with $E$ at smaller $T$.

\begin{figure}[ht]
\begin{center}
\includegraphics{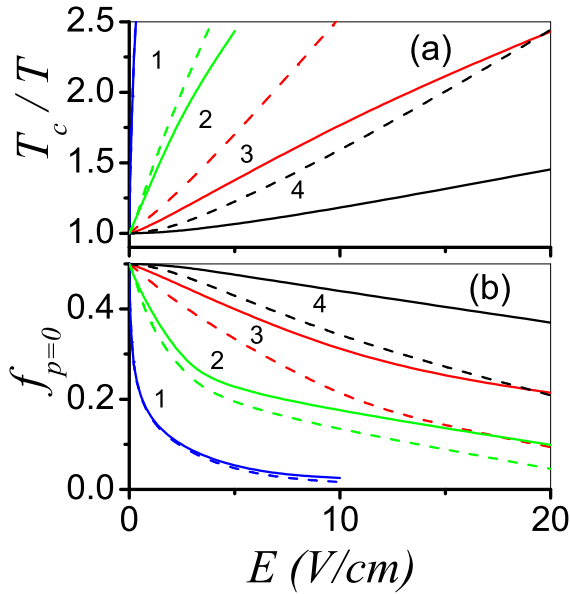}
\end{center}
\addvspace{-1 cm} \caption{(Color online) Determined from the
balance Eqs. (\ref{16}), (\ref{17}) dimensionless effective
temperature $T_c/T$ (a) and maximal distribution $\tilde{f}_{p=0}$
(b) versus electric field for $T=$20 K (1), 77 K (2), 150 K (3),
and 300 K (4). Solid and dashed curves are correspondent to
$l_c=$10 nm and 20 nm, respectively.}
\end{figure}

\subsection{Hot carrier concentration}
Using the solutions of Eq. (\ref{15}) or of the balance Eqs.
(\ref{16})-(\ref{18}), we calculate below the sheet concentration
which is determined as follows
      \begin{equation}
      n=\frac{2}{\pi\hbar^2}\int\limits_0^\infty dp p f_{p} . \label{14}
      \end{equation}
Here in the right hand side $f_{p}$ is replaced by
$\widetilde{f}_p$ for the case $II$. As a result, the field
dependence of $n$ is determined by a competition of the effective
temperature $T_{c}$ grows and the maximal distribution
$\tilde{f}_{p=0}$ decrease, see Figs. 3a and 3b: so that the
concentration will grow slowly with the field.

\begin{figure}[ht]
\begin{center}
\includegraphics{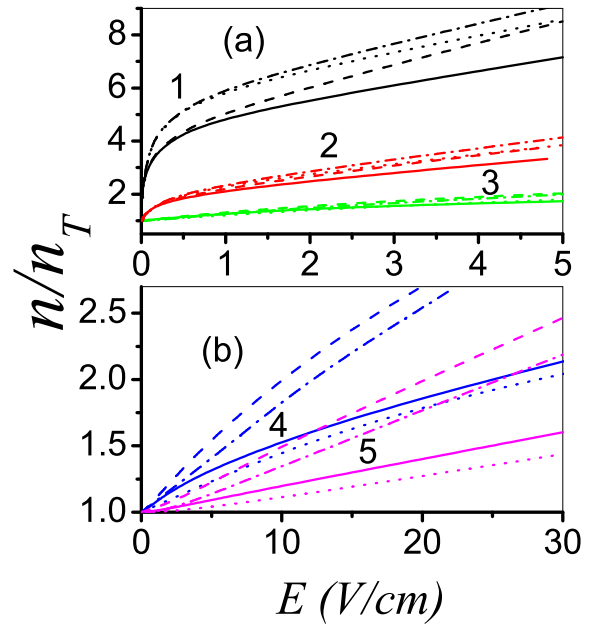}
\end{center}
\addvspace{-1 cm} \caption{(Color online) Normalized carrier
concentration, $n/n_{T}$, versus electric field: (a) at $T=$4.2 K
(1), 20 K (2), 77 K (3) and (b) at $T=$150 K (4), 300 K (5). The
solid and the dashed curves are calculated from Eq. (\ref{15}) for
$l_c=$10 nm and 20 nm, respectively. The dotted and the dot-dashed
curves are calculated from Eqs. (\ref{16})-(\ref{18}) for $l_c=$10
nm and 20 nm, respectively.}
\end{figure}

In Fig. 4 we plot the dimensionless carrier concentrations
$n/n_T$ as function of $E$. Here the equilibrium concentration readily follows
from Eq. (\ref{14}) as $n_T=\pi (T/\hbar v_W)^{2}/6$ [notice, at $T=4.2$ K,
20 K, 77 K, 150 K, and 300 K one obtains $n_{T}=1.6 \times 10^{7} cm^{-2}$, $3.6
\times 10^{8} cm^{-2}$, $5.4 \times 10^{9} cm^{-2}$, $2.0 \times
10^{10} cm^{-2}$, and $8.1 \times 10^{10} cm^{-2}$, respectively].
From Fig. 4 it is seen that: i) at low temperatures (and not too
large $E$) the relative increase of the density $(n-n_T)/n_T$ is
approximately $\propto E^{1/2}$ as due to small characteristic
momentum, $\overline{p}$, effect of a finite $l_c$ is negligible;
ii) at higher temperatures (or very large $E$) $(n-n_T)/n_T
\approx A+B E$, i.e., it is linear as due to large $\overline{p}$
effect of a finite $l_c$ becomes essential. Notice, that the
density grows becomes faster for larger $l_c$.

\begin{figure}[ht]
      \begin{center}
\includegraphics{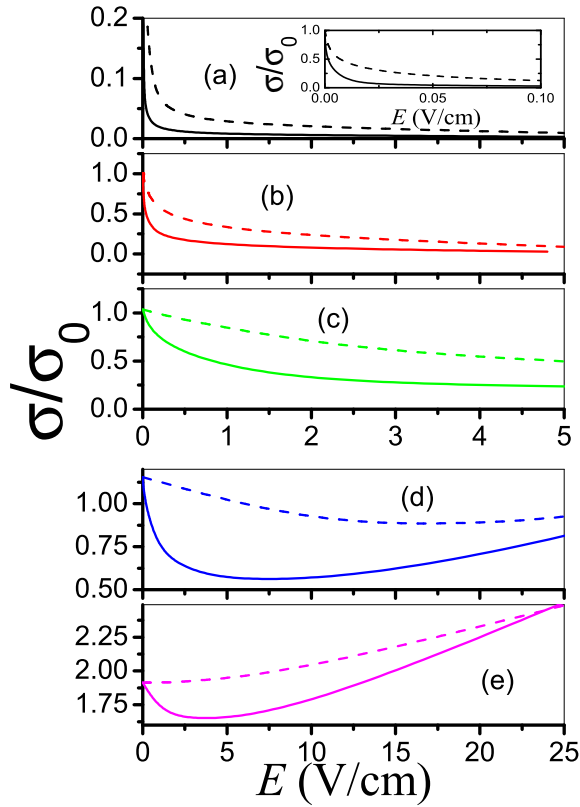}
      \end{center}\addvspace{-1 cm}
\caption{Dimensionless conductivity, $\sigma/\sigma_{0}$, versus field $E$
for $l_c=$10 nm at different temperatures: $T=4.2$ K (a), 20 K (b), 77 K (c),
150 K (d), and 300 K (e). The solid and the dashed curves are calculated
from Eqs. (\ref{15}), (\ref{19}) and Eqs. (\ref{16})-(\ref{18}), (\ref{19}),
respectively. Inset in panel (a) shows low-field dependencies.}
\end{figure}

\begin{figure}[ht]
      \begin{center}
\includegraphics{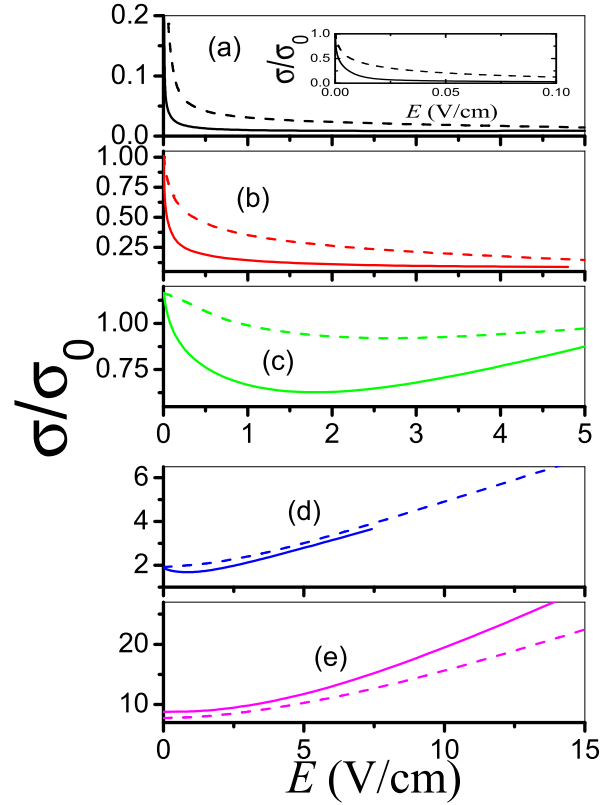}
      \end{center}\addvspace{-1 cm}
\caption{The same as in Fig. 5 for $l_c=$20 nm. }
\end{figure}

\section{Current-voltage characteristics}
Using obtained in Secs. IIIA and IIIB nonequilibrium distribution
functions we calculate here nonlinear conductivity introduced by Eq.
(\ref{6}) and analyze modifications of the current voltage
characteristics due to temperature and $l_{c}$ variations. Performing the
integration in Eq. (\ref{6}) by parts one arrives to
      \begin{equation}
      \sigma=\sigma_{0}\left[2f_{p=0}+\frac{l_c}{\hbar}\int\limits_0^{\infty}
      dp f_p \Phi(pl_c/\hbar )\right] ,
      \label{19}
      \end{equation}
where $\sigma_{0}=(2v_W/v_d)e^{2}/\pi\hbar$ is the characteristic
conductivity. For the case of short-range scattering, $\overline{p}l_c
/\hbar\ll 1$ ($\overline{p}$ is the characteristic momentum of hot carriers)
      conductivity can be expressed through the distribution function of
      low energy carriers as: $\sigma \simeq 2\sigma_{0}f_{p=0}$. As $f_{p=0}=1/2$ for
      $E \to 0$, it follows that $\sigma_{0}$ is the linear (in the absence of heating)
conductivity in short-range scattering limit. In Fig. 5 we plot
$\sigma/\sigma_{0}$ versus field $E$ for $l_c$=10 nm. Notice, in
Figs. 5a-5c, except the inset of Fig. 5a, the same field region is
used; it is different from the one used in Figs. 5d-5e. It is seen
that both approaches give similar dependences even for very strong
field, for given $T$ and $l_c$. The dependencies
$\sigma/\sigma_{0}$ vs $E$ for $l_c$=20 nm are plotted in Fig. 6,
where conductivity grows faster for higher temperatures.

Next, we turn to consideration of the current-voltage
characteristics (i.e., the current density $I=\sigma E$ versus
field $E$).  In Fig. 7 we plot the current density, $I=\sigma E$,
versus field $E$ for $l_c$=10 nm at different temperatures.
Similar dependencies for $l_c$=20 nm are shown in Fig. 8. It is
seen that current-voltage characteristics can be essentially
nonlinear already for relatively small $E$ and $I$ starting from
$10^{-5}$ A/cm in the low-temperature region. Point out that for
larger $E$ and $I$ (if $E \geq 1$ V/cm at low temperatures and $E
\geq 10$ V/cm at room temperature) Figs. 7 and 8 will present
dependences that are close to linear ones, $A+B E$. In addition,
Figs. 7 and 8 show that from both approaches (cases $I$ and $II$)
similar current-voltage characteristics are obtained, for given
$T$ and $l_c$. It is also seen that current-voltage
characteristics can manifest as sublinear so superlinear form in
wide $E$ regions depending on value of $T$, $l_c$; these
modifications of the form are essentially related with the effects
due to finite $l_c$. 

\begin{figure}[ht]
      \begin{center}
\includegraphics{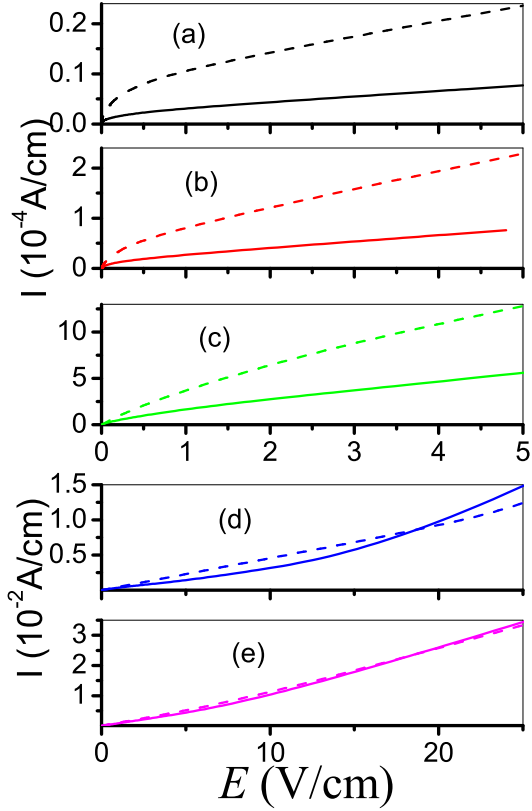}
      \end{center}\addvspace{-1 cm}
      \caption{Current-voltage characteristics for $l_c=$10 nm at
different temperatures: $T=4.2$ K (a), 20 K (b), 77 K (c), 150 K (d),
and 300 K (e). The solid and the dashed curves are calculated
from Eqs. (\ref{15}), (\ref{19}) and Eqs. (\ref{16})-(\ref{18}), (\ref{19}),
respectively.}
      \end{figure}

\begin{figure}[ht]
      \begin{center}
\includegraphics{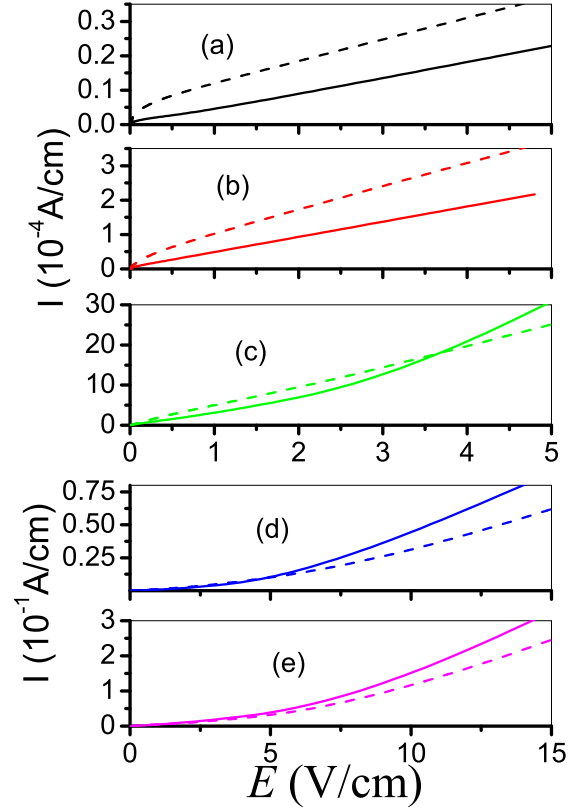}
      \end{center}\addvspace{-1 cm}
\caption{The same as in Fig. 7 for $l_c=$20 nm. }
\end{figure}
      \section{Conclusions}
To summarize, we developed the theory of the carries heating in intrinsic
graphene under strong dc electric field for the cases when intercarrier
scattering is negligible or dominant. It is found that the
      deviation from the equilibrium distribution starts from very low
      electric fields (from $0.3$ V/cm for the room temperature and from
      10$~\mu$V/cm for the liquid helium temperature; in particular,
      for a finite fields $df_p/dp$ tends to zero for $p \ll p_E \alt p_T$,
      (while for equilibrium distribution $df_{p}^{eq}/dp \approx
      -1/4p_T$ has the finite constant value) due to ineffective
      energy relaxation of slow carriers. Since with the increase of
      heating this relaxation sharply grows and the recombination of
      carriers is dominant over the thermogeneration within the high
      energy region, it follows that the current-voltage characteristics
      should be sublinear for the short-range scattering case. The
superlinear dependency that is realized in strong fields for
scattering on finite-range disorder. Therefore, for intrinsic
graphene we obtain unusual combination of the low threshold of
nonlinearity and appearance of the second ohmic region for strong
fields.

      Let us discuss the main assumptions made. Here it is studied
      intrinsic graphene which has at most resistivity so the effect of
      heating manifests itself the most. Doped materials
      require of separate investigation, moreover nonlinearity in these
      materials must be weaker. We have restricted ourselves by study only
      of limiting cases of the absence of intercarrier scattering
      or dominating Coulomb scattering that imposes
      quasiequilibrium distribution. Point out, that the field
      dependences of concentration and the current-voltage characteristics
      there are in good agreement (the strict solution of the problem should
be somewhere between the solutions obtained), i.e., they
      have weak sensitivity to the details of distribution function. As
      the main generation-recombination mechanism it is assumed
      radiative-induced interband transitions because Auger-processes
      are forbidden due to the symmetry of electron-hole states
      \cite{14}. Possible contribution of other generation-recombination
      mechanisms (e.g., disorder induced transitions, caused by acoustic
      phonons or carrier-carrier scattering) demands additional
      investigation. At last, we consider interaction of carriers with
      equilibrium thermostat formed by acoustic phonons and radiation
(compare with Ref. 18). The problem of heat removal is beyond of the scope
of this paper. We speculate now that heat removal is sufficiently effective
      \cite{7}. In addition, it is possible to use short pulses (all
      times of relaxation there are $< 1~\mu$s), when the thermostat
      will not be overheated. These limitations are related with the
      lack of data on graphene, however, the qualitative picture will
      not essentially modify after taking into account with better
      precision of the parameters and the mechanisms of relaxation.

      The rest of assumptions there are standard. To describe the
      momentum relaxation it is taken into account only the statical
disorder scattering using the phenomenological model of Ref. 11 (despite
that, it corresponds well the experimental data, the microscopic
mechanisms of scattering are still unclear \cite{19}) and small contribution
of acoustic phonons is discarded. Using of the quasielastic approximation
for describing of the energy relaxation is justified by condition $v_W
      \gg s$, that ensures the small energy transference for the
      scattering process. As the energy of optical phonons is large,
      their contribution can be neglected even in considered strong
      fields. Also it is unimportant in considered range of the energies
      an anisotropy of both electron and phonon spectra. Interaction
      with thermic radiation can be described by taking into account
      only of direct interband transitions (the Drude absorption is
      small). Therefore, applied approximations accurately describe the
      heating mechanism in graphene and lead to correct quantitative
      estimation of current-voltage characteristics.

      In conclusion, prospects of many device applications, in
      particular, the characteristics of field-effect transistors or
      effectiveness of graphene interconnections, are substantially
      dependent on the carries heating. In addition, study of hot electrons
      (holes) gives information on electron-phonon coupling and about
      the mechanisms of recombination, i.e., makes possible verification
      of the relaxation mechanisms. Therefore, we believe that the obtained
      results  will stimulate further experimental and theoretical study
      (as well as numerical modeling) of hot carriers in graphene.

      \appendix*
      \section{Kinetic equation}
      Below we evaluate the system of quasiclassical kinetic equations
      governing the nonequilibrium carrier distributions in graphene placed in a
      strong electric field $\bf E$. Such a consideration is analogous
      to the approach used for the bulk narrow-gap semiconductors, see
      \cite{20}) and the recent papers \cite{10} which develop a similar
      approach for the case of graphene. But the case of the bipolar
      electron-hole plasma is not analyzed in these papers.

We start from the single-particle density matrix which is governed by the
      quantum kinetic equation \cite{15}:
      \begin{equation}
      \frac{i}{\hbar }\left[\hat h_W-e{\bf E}\cdot{\bf
      x},\hat\rho\right] =\widehat J_{coll} .
      \end{equation}
      The collision integral $\widehat J_{coll}$ represents studied
      above scattering mechanisms and $\hat h_W$ is the $2\times 2$
      Weyl-Wallash Hamiltonian, \cite{21} describes states nearby the band
      cross-point. Here we use the linear dispersion laws
      $\varepsilon_{lp}= lv_{W}p$, with $l$ corresponding to the
      conduction ($l$=+1) or the valence ($l$=-1) bands, the
      eigenvectors $|l{\bf p}\rangle$ are defined from the eigenstate
      problem:
      \begin{eqnarray}
      \hat{h}_{W}|l{\bf p}\rangle =\varepsilon_{lp}|l{\bf p}\rangle
      , ~~~~ \hat{h}_{W}=v_{W}(\hat{\bsigma}\cdot{\bf p}), ~~~~
      \\ |+1{\bf p}\rangle =\frac{1}{\sqrt{2}}\left|\begin{array}{l} ~1
      \\ e^{i\phi} \end{array} \right| , ~~~~~|-1{\bf p}\rangle
      =\frac{1}{\sqrt{2}}\left|\begin{array}{l} -e^{-i\phi}
      \\ ~~~ 1 \end{array} \right| , \nonumber
      \end{eqnarray}
      where $\phi$ is the $\bf p$-plane polar angle.
      Distribution function over $l{\bf p}$ states $F_{l{\bf
      p}}=\left\langle {l{\bf p}|\hat \rho |l{\bf p}} \right\rangle$ and
      the non-diagonal part of density matrix $\widetilde F_{\bf p}  =
      \left\langle {1{\bf p}|\hat \rho | - 1{\bf p}} \right\rangle =
      \left\langle { - 1{\bf p}|\hat \rho |1{\bf p}} \right\rangle^*$
      are obtained from the system of kinetic equations:
      \begin{eqnarray}
      e{\bf E}\cdot\frac{\partial F_{l{\bf p}}}{\partial{\bf
      p}}+l\frac{e}{\hbar} {\bf E}\cdot\left\{ {\widetilde F_{\bf p}
      {\bf X}_{-l,l} ({\bf p}) - {\bf X}_{l,-l} ({\bf p})\widetilde
      F_{\bf p}^*} \right\} \nonumber \\
      =\left\langle l{\bf p}\left|J_c(\hat\rho )\right|l{\bf p}
      \right\rangle~~~~~~~~~~~~~~~~  \\
      \frac{i}{\hbar}2v_{W}p\widetilde F_{\bf p}+e{\bf
      E}\cdot\frac{\partial\widetilde F_{\bf p}} {\partial {\bf
      p}}+\frac{e}{\hbar }{\bf E}\cdot{\bf X}_{1,-1}({\bf p}) \nonumber \\
      \times\left( {F_{1{\bf p}}-F_{-1{\bf p}} } \right)  + \frac{e}{\hbar
      }{\bf E} \cdot \left\{ {\bf X}_{-1,-1}({\bf p})-{\bf X}_{1,1}
      ({\bf p})\right\} \widetilde F_{\bf p} \nonumber \\
      =\left\langle +1{\bf p}\left| J_c(\hat\rho )\right| -1{\bf p}
      \right\rangle  ~~~~~~~~~~~~~~~~
      \end{eqnarray}
      Here the interband matrix element of coordinate, ${\bf
      X}_{l,l'}({\bf p})= \left\langle {l{\bf p}|\hat x|l{\bf p}}
      \right\rangle$, is calculated on wave functions of the momentum
      representation  (A.2) and has the value  of the order of $\hbar
      /\bar p$, where $\bar p$ is the characteristic momentum of the hot
      carries.

In order to show the smallness of nondiagonal components
      of the distribution function, the estimation $\left\langle l{\bf
      p}\left| J_c(\hat\rho ) \right|l{\bf p} \right\rangle\sim F_{l{\bf
      p}}/\tau_m$ is also used, where $\tau_m$ is the characteristic
      time of the momentum relaxation (the smallest characteristic time
of the problem). For typical conditions of the applicability of quasiclassical
      description \cite{13}
      \begin{equation}
      2v_W\bar p\gg\frac{\hbar}{\tau _c}, ~~~~ eE\tau_c\ll \bar p
      \end{equation}
      from Eq. (A.4) we have estimation $\widetilde F_{\bf p} /
      F_{l{\bf p}}\sim \hbar /(2v_W\bar p \tau_c)$.

Further, neglecting by small nondiagonal contributions we arrive to the
usual system of kinetic equations:
      \begin{equation}\label{eq1}
      e{\bf E}\cdot\frac{\partial F_{l{\bf p}}}{\partial {\bf p}}=
      \sum_jJ_j\{F|l{\bf p}\},
      \end{equation}
      which describes the distribution of hot carriers over the bands
      $l=\pm 1$ and the 2D momentum $\bf p$. The scattering integrals,
      $J_j\{F|l{\bf p}\}$, are obtained  in Ref. \cite{5,10}.
      Here $l$ corresponds to conduction ($l$=+1) or valence ($l$=-1)
      band, ${\bf p}$ is the 2D momentum, $J_j\{F|l{\bf p}\}$ is the
      collision integral for the $j$th scattering mechanism, [$j=D,~LA,~R$,
      and $C$, see discussion after Eq.(1)]. It is convenient to make
transition to the electron-hole representation introducing the electron
($e$) and hole ($h$) distribution functions, $f_{e{\bf p}}$ and
$f_{e{\bf p}}$, according to the replacements \cite{22}:
      \begin{equation}
      F_{+1,{\bf p}}\rightarrow f_{e{\bf p}}, \qquad 1-F_{-1,{\bf p}t}
      \rightarrow f_{h{\bf p}}
      \end{equation}
and to rewrite the collision integrals in (A.6) through $f_{e,h{\bf p}}$.

Finally, using the substitution (A.7) and the velocity operator
$\hat{\bf v}=i[\hat{h}_W,{\bf x}]/ \hbar =v_W\hat{\bsigma}$, we
      obtain the velocity of $l{\bf p}$ state: $\left\langle l{\bf p}
      |\hat{\bf v}|l{\bf p}\right\rangle=lv_W{\bf p}/p\equiv{\bf
      v}_{l{\bf p}}$. Taking into account four-fold degeneracy of states in
      graphene due to spin and valley degrees of freedom, one obtains
      the current density
      \begin{equation}
      {\bf I}=\frac{4e}{L^2}\sum_{l{\bf p}}{\bf v}_{l{\bf p}}F_{l{\bf
      p}}=\frac{4ev_W}{L^2} \sum_{\bf p}\frac{\bf p}{p}\left(
      f_{e{\bf p}}+f_{h{\bf p}}\right) .
      \end{equation}
      In addition, the intrinsic material the electron density is equal to the
      hole one (the neutrality condition):
      \begin{equation}
      \frac{4}{L^2}\sum_{\mathbf{p}}\left(
      f_{e\mathbf{p}}-f_{h\mathbf{p}}\right) =0 .
      \end{equation}
      Thus, the symmetric scattering for the $c$- and the
      $v$- bands Eqs. (A6), (A8) and (A9) preserve their form when
      $f_{e{\bf p}}$ is replaced by $f_{h{\bf p}}$. Thus, the electron and hole
      distributions in the intrinsic material are identical,
      $f_{e{\bf p}}=f_{h{\bf p}}\equiv f_{\bf p}$, so that the kinetic equation and the
      current density take forms (1) and (3), respectively.


\begin{thebibliography}{52}
\bibitem{16}
      The relation $A_p+p(dA_p/dp)/2=[d(p^2 A_p)/dp]/2p$ is used in Eq. (8).
\end{thebibliography}
      \end{document}